\documentclass[preprint, superscriptaddress, showpacs,preprintnumbers,amsmath,amssymb]{revtex4}
\usepackage{graphicx}

\begin{document}

\thispagestyle{empty}

\title{New constraints on the Yukawa-type corrections to Newtonian
gravity at  short separations}

\author{
G.~L.~Klimchitskaya
}
\affiliation{Central Astronomical Observatory
at Pulkovo of the Russian Academy of Sciences,
St.Petersburg, 196140, Russia}
\affiliation{Institute of Physics, Nanotechnology and
Telecommunications, St.Petersburg State
Polytechnical University, St.Petersburg, 195251, Russia}
\affiliation{E-mail: g.klimchitskaya@gmail.com}
\author{
 V.~M.~Mostepanenko}
\affiliation{Central Astronomical Observatory
at Pulkovo of the Russian Academy of Sciences,
St.Petersburg, 196140, Russia}
\affiliation{Institute of Physics, Nanotechnology and
Telecommunications, St.Petersburg State
Polytechnical University, St.Petersburg, 195251, Russia}
\affiliation{E-mail: vmostepa@gmail.com}

\begin{abstract}
We discuss the strongest constraints on
the Yukawa-type corrections to Newton's gravitational law
within a submicrometer interaction range following
from measurements of the Casimir force.
In this connection the complicated problems arising when
comparing the measurement data with the Lifshitz theory
are analyzed. Special attention is paid to the results of
two recent experiments on measuring the Casimir interaction
between ferromagnetic surfaces and sinusoidally corrugated
surfaces at various angles between corrugations.
\end{abstract}
\pacs{14.80.-j, 04.50.-h, 04.80.Cc, 12.20.Fv}

\maketitle
\section{Introduction}

Among all fundamental interactions gravitation is the most
commonly known and simultaneously the most difficult for both
experimental and theoretical investigation. It might be
considered paradoxical that up to the present the gravitational
constant is measured with less precision than other fundamental
physical constants. During the last centure research in
gravitation was somewhat isolated from all other branches of
physics dealing mostly with quantum phenomena.
Many attempts were undertaken to combine gravitation with other
interactions in the framework of some unified description, e.g.,
supergravity, but all of them till the moment are only impressive
mathematical schemes rather than successful physical theories.
Against this background a few achievements presenting consistent
and physically reasonable unification between gravitation and
quantum phenomena in some special cases are of even greater
value.

One of these achievements is the quantum field theory in
spatially homogeneous isotropic space-time developed by
Prof.\ A.\ A.\ Grib and his collaborators (see
papers \cite{1,2,3,4,5,6}, review \cite{7} and monograph
\cite{8}). This theory was applied to the Friedmann
cosmological models describing our Universe and found a lot of
prospective applications to the effects of particle creation
from vacuum by the nonstationary gravitational field,
polarization of vacuum and spontaneous symmetry breaking.
On similar grounds Prof.\ A.\ A.\ Grib and his collaborators
developed the theory of particle creation by a nonstationary
electric field \cite{9,10} (recently the same methods were
applied \cite{11} to describe the creation of quasiparticles
in graphene).

Taking into account the lack of experimental information on
the border between gravitational physics and quantum phenomena,
much attention has been recently paid to the search of
Yukawa-type corrections  to Newtonian gravity at short
separations \cite{12}. Such corrections arise due to exchange
of hypothetical light elementary particles predicted \cite{13}
by many extensions of the Standard Model and in extra-dimensional
physics with low-energy compactification scale \cite{14}.
In the range of separations between the test bodies below a few
micrometers Newton's law of gravitation is not verified
experimentally, so that corrections to it are possible which
exceed gravity by many orders of magnitude. These corrections
cannot be constrained with the help of standard gravitational
experiments of E\"{o}tvos and Cavendish type. The point is that
at so small separations the van der Waals and Casimir forces
which act between the closely spaced surfaces of probe masses
due to electromagnetic fluctuations become much larger than the
gravitational force. The latter loses its sensitivity to the
presence of possible corrections. Because of this in the
interaction range below a few micrometers it was proposed
\cite{15,15a} to use measurements of the van der Waals and
Casimir force for obtaining stronger constraints on the
corrections to Newton's gravitational law.

During the last 15 years  a lot of experiments on measuring
the Casimir force between metallic, semiconductor and dielectric
test bodies has been performed \cite{16,17}. As a result,
the previously known constraints on the Yukawa-type corrections
to Newtonian gravity in the submicrometer interaction range were
strengthened up to a factor of 24 millions \cite{18}.
In doing so unexpected theoretical problems related to the
comparison between experiment and theory have been analyzed.
In the present paper we briefly summarize the strongest
constraints on the Yukawa-type corrections
to Newtonian gravity following from measurements of the Casimir
force. We discuss the reliability of these constraints in
connection with abovementioned problems arising in the
fluctuational electrodynamics. We also present the most recent
constraints obtained from two experiments performed in 2013.

The paper is organized as follows. In Sec.~II we introduce the
used notations and parametrizations and summarize the results
obtained in the past. Section~III is devoted to the comparison
between measurements of the Casimir force and theory.
In Sec.~IV we present the constraints on corrections to
Newton's law obtained from two most recent experiments on
measuring the Casimir force between ferromagnetic and
corrugated surfaces. Section~V is devoted to our conclusions
and discussion.

\section{Yukawa-type corrections to Newton's gravitational law
and constraints on them from measurements of the Casimir force}

It is conventional to present the gravitational potential between
the two pointlike masses $m_1$ and $m_2$ spaced at a separation
$r$ as a sum of the Newtonian part $V_N(r)$ and the Yukawa-type
correction $V_{\rm Yu}(r)$:
\begin{equation}
V(r)=V_N(r)+V_{\rm Yu}(r)=-\frac{Gm_1m_2}{r}\left(
1+\alpha e^{-r/\lambda}\right).
\label{eq1}
\end{equation}
\noindent
Here, $G$ is the Newtonian
gravitational constant, and $\alpha$ and $\lambda$ are
the strength and interaction range of the Yukawa-type
correction.
If the Yukawa-type correction to Newtonian gravitational
potential $V_{N}(r)$ is caused by an exchange of light
bosons of mass $m$ between the probe masses $m_1$ and $m_2$,
the interaction range $\lambda$ has the meaning of the
Compton wavelength of this boson $\lambda=\hbar/(mc)$.
Alternatively, if the Yukawa-type correction arises due
to the compactification of extra spatial dimensions in
multi dimensional schemes, the quantity $\lambda$ has
the physical meaning of the size of the compact manifold.

As was mentioned in Sec.~I, at separations below a few
micrometers the Newtonian gravitational force becomes smaller
than the van der Waals and Casimir forces acting between
closely spaced surfaces. In fact it is even smaller than the
error in the measurements of the van der Waals forces.
Because of this, when calculating the interaction energy of two
macroscopic bodies due to potential (\ref{eq1}), one can
neglect by the Newtonian contribution and integrate the
Yukawa-type correction alone over the volumes of both bodies
\begin{equation}
V_{\rm Yu}(a)=-G\alpha\int_{V_1}d^3r_1
\rho_1(\mbox{\boldmath$r$}_1)\int_{V_2}d^3r_2
\rho_2(\mbox{\boldmath$r$}_2)
\frac{e^{-|{\scriptsize{\mbox{\boldmath$r$}_1-
\mbox{\boldmath$r$}_2}}|/\lambda}}{|\mbox{\boldmath$r$}_1-
\mbox{\boldmath$r$}_2|}.
\label{eq2}
\end{equation}
\noindent
Here, $\rho_1(\mbox{\boldmath$r$}_1)$ and
$\rho_2(\mbox{\boldmath$r$}_2)$ are the mass densities
of, generally speaking, nonhomogeneous test bodies and
 $a$ is the closest separation between them.
Then the Yukawa-type force and its gradient
are given by
\begin{equation}
F_{\rm Yu}(a)=-\frac{\partial V_{\rm Yu}(a)}{\partial a},
\qquad
\frac{\partial F_{\rm Yu}(a)}{\partial a}
=-\frac{\partial^2 V_{\rm Yu}(a)}{\partial a^2}.
\label{eq3}
\end{equation}

In the experiments on measuring the Casimir force (see
reviews in Refs.~\cite{16,17}) either the force
$F_C(a,T)$ acting between two test bodies or its gradient
$\partial F_C(a,T)/\partial a$ have been measured
($T$ is the temperature at which measurements under
consideration are performed). The measurement results were
compared with theoretical predictions for the Casimir
force and its gradient and good agreement was found in the
limits of some experimental errors $\Delta_{F_C}(a)$ and
$\Delta_{F_C^{\prime}}(a)$, respectively.
Thus, within the limits of these errors no hypothetical
Yukawa-type interaction was observed. The respective
constraints on the parameters of Yukawa interaction
$\alpha,\,\lambda$ follow from the inequalities
\begin{equation}
|F_{\rm Yu}(a)|\leq\Delta_{F_C}(a),
\quad
\left|\frac{\partial F_{\rm Yu}(a)}{\partial a}\right|
\leq\Delta_{F_C^{\prime}}(a),
\label{eq4}
\end{equation}
\noindent
where $\alpha$- and $\lambda$-dependent expression for
$F_{\rm Yu}(a)$ is given by Eqs.~(\ref{eq2}) and (\ref{eq3}).

\begin{figure}[b]
\vspace*{-12cm}
\centerline{\hspace*{3cm}
\includegraphics{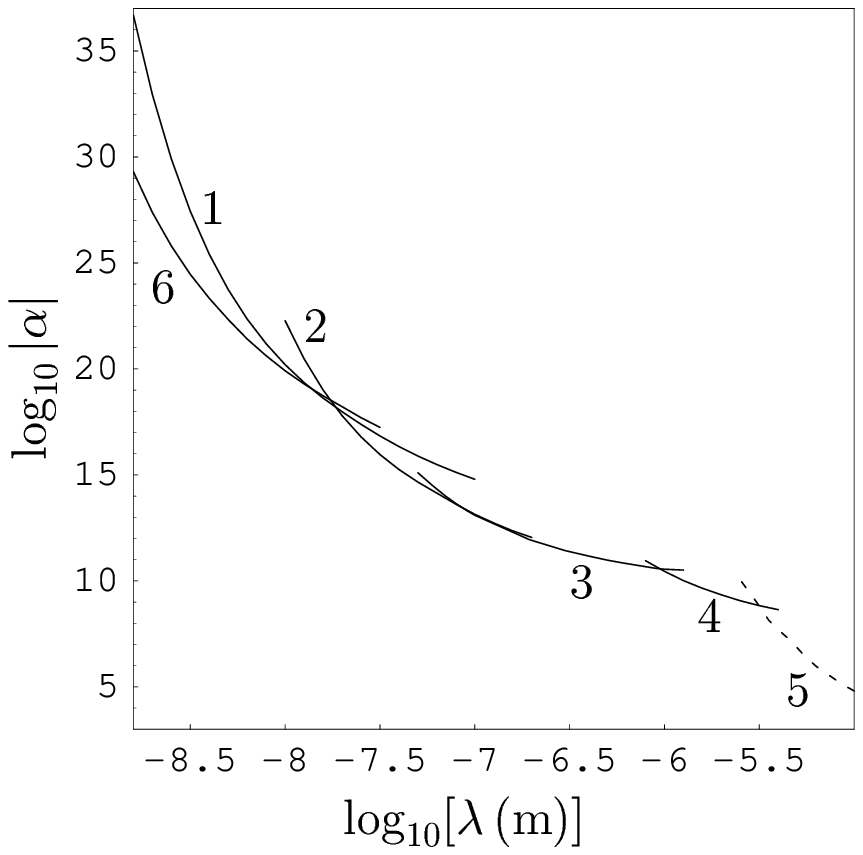}
}
\vspace*{-7cm}
\caption{\label{fg1}
Constraints on the  Yukawa-type corrections to
Newton's gravity from different experiments are shown by
the lines 1--6 (see  text for further discussion).
The region of $(\lambda,\alpha)$ plane above each line
is prohibited and below each line is allowed.
}
\end{figure}

Now we list the most strong constraints on the parameters
$\alpha$ and $\lambda$ obtained from measurements of the
Casimir force performed before 2013. The constraints on
$\alpha$ and $\lambda$ are usually presented as some lines
in the ($\lambda,\,\alpha$) plane where the region of this
plane above the line is prohibited by the results of respective
experiment and the region below the line is allowed.
By line~1 in Fig.~\ref{fg1} we show constraints obtained
\cite{19,20} from measurements of the Casimir force between
Au-coated surfaces of a microsphere and a plate by means
of an atomic force microscope \cite{21}. Line~2 in the same
figure shows constraints obtained from measuring the
gradient of the Casimir force between similar surfaces by
means of a micromachined oscillator \cite{19,22}.
Line~3 in Fig.~\ref{fg1} follows from the so-called
Casimir-less experiment where the contribution of the
Casimir force acting between a microsphere and a plate was
compensated using some special arrangement of the setup
\cite{23}. Finally, line~4 shows the constraints obtained
\cite{24} from measurements of the Casimir force between
Au-coated surfaces of a plate and a spherical lens of
centimeter-size radius of curvature by means of torsion
pendulum.

As can be seen in Fig.~\ref{fg1}, the strength of constraints
following from the Casimir effect quickly increases with the
increase of $\lambda$. However, for $\lambda$ exceeding
several micrometers the strongest constraints on
$\alpha$ and $\lambda$ follow not from measurements of the
Casimir force but from gravitational experiments.
To illustrate this, in Fig.~\ref{fg1} we plot by the line~5
the constraints obtained from the most precise Cavendish-type
experiment of Refs.~\cite{25,26}. At the same time, with
decreasing $\lambda$ down to 1\,nm the strength of
constraints shown by the lines 1--4 quickly decreases.
It was shown \cite{18} that within the interaction range
from 1.6 to 14\,nm the strongest constraints on
$\alpha,\,\lambda$ follow from measurements of the lateral
Casimir force which arises between sinusoidally corrugated
surfaces of a sphere and a plate with common period
\cite{27,28}. These constraints are shown by the
line~6 in Fig.~\ref{fg1}. At even shorter $\lambda$ below
1\,nm the strongest constraints on the Yukawa-type corrections
to Newtonian gravity follow from precision atomic
physics \cite{29}.

In the end of this section it is worth noting that in
Fig.~\ref{fg1} we do not show constraints obtained from
measurements of the Casimir force with the help of torsion
pendulum presented in Refs.~\cite{30,31,32,33}.
The point is that these are not direct measurements of the
Casimir force, but of much larger force of unknown nature
from which the Casimir contribution was extracted by means
of the fitting procedure using some postulated theoretical
expressions. The critical discussion of these experiments
contained in the literature \cite{16,17,34,35,36,37,38}
leads to a conclusion that both the measured data and respective
constraints are not reliable.

\section{Problems in experiment-theory comparison for
the Casimir force}

As explained in Sec.~II, the strongest constraints on the
Yukawa-type corrections to Newtonian gravity in the
submicrometer interaction range are obtained from the measure
of agreement between the experimental data and theoretically
calculated Casimir forces. Because of this, both the solid
data and consistent theory are required for the reliability
of these constraints. The fundamental theory of the van der Waals
and Casimir forces used in calculations was developed by
Lifshitz \cite{39} in the framework of fluctuational
electrodynamics and is commonly known as the Lifshitz theory.
In this theory the Casimir free energy ${\cal F}_C(a,T)$ and
force ${F}_C(a,T)$ are expressed via the frequency-dependent
dielectric permittivities of the interacting bodies.
For Au bodies used in most of experiments the complex index
of refraction $n(\omega)$ [and respective dielectric
permittivity $\varepsilon(\omega)]$ is measured over a wide
range of frequencies. The dielectric permittivity at very
low frequencies beyond this range (which is also needed in
computations using the Lifshitz theory) is obtained by means
of extrapolation of the measured optical data with the help
of well tested Drude model
\begin{equation}
\varepsilon_D(\omega)=1-
\frac{\omega_p^2}{\omega[\omega+i\gamma(T)]}.
\label{eq5}
\end{equation}
\noindent
Here $\omega_p$ is the plasma frequency and
$\gamma(T)\ll\omega_p$ is the relaxation parameter.
Equation (\ref{eq5}) demonstrates that at very low,
quasistatic, frequencies $\omega\ll\gamma(T)$ the dielectric
permittivity behaves as $\varepsilon_D(\omega)\sim 1/\omega$,
as it must be in accordance with the Maxwell equations.

The first unexpected problem arising in the Lifshitz theory
is that it violates the third law of thermodynamics (the
Nernst heat theorem) when the interacting bodies are described
by the dielectric permittivity (\ref{eq5}) \cite{40,41,42}.
Specifically, it was shown \cite{40,41,42} that the Casimir
entropy
\begin{equation}
S_C(a,T)=-\frac{\partial{\cal F}_C(a,T)}{\partial T}
\label{eq6}
\end{equation}
\noindent
goes to a nonzero negative limit depending on the parameters of
the system when the temperature vanishes.
It was shown also \cite{40,41,42} that the violation disappears
when one neglects by the relaxation, i.e., suggests that
$\gamma(T)=0$. In this case the dielectric permittivity is
described by the so-called plasma model
\begin{equation}
\varepsilon_p(\omega)=1-
\frac{\omega_p^2}{\omega^2},
\label{eq7}
\end{equation}
\noindent
which in fact valid only in the region of very high frequencies
$\omega\gg\gamma(T)$ characteristic for infrared optics.
Keeping in mind that the fulfilment of the Nernst heat theorem
is caused by the low-frequency behavior of $\varepsilon$,
where not Eq.~(\ref{eq7}) but Eq.~(\ref{eq5}) is correct,
the above facts should be considered as somewhat paradoxical.
The discussion of this subject can be found, e.g., in
Refs.~\cite{43,44}. It was even suggested \cite{45} that there
might be profound difference in the reaction of a physical system
to the real and fluctuating electromagnetic fields.

\begin{figure}[b]
\vspace*{-12cm}
\centerline{\hspace*{3cm}
\includegraphics{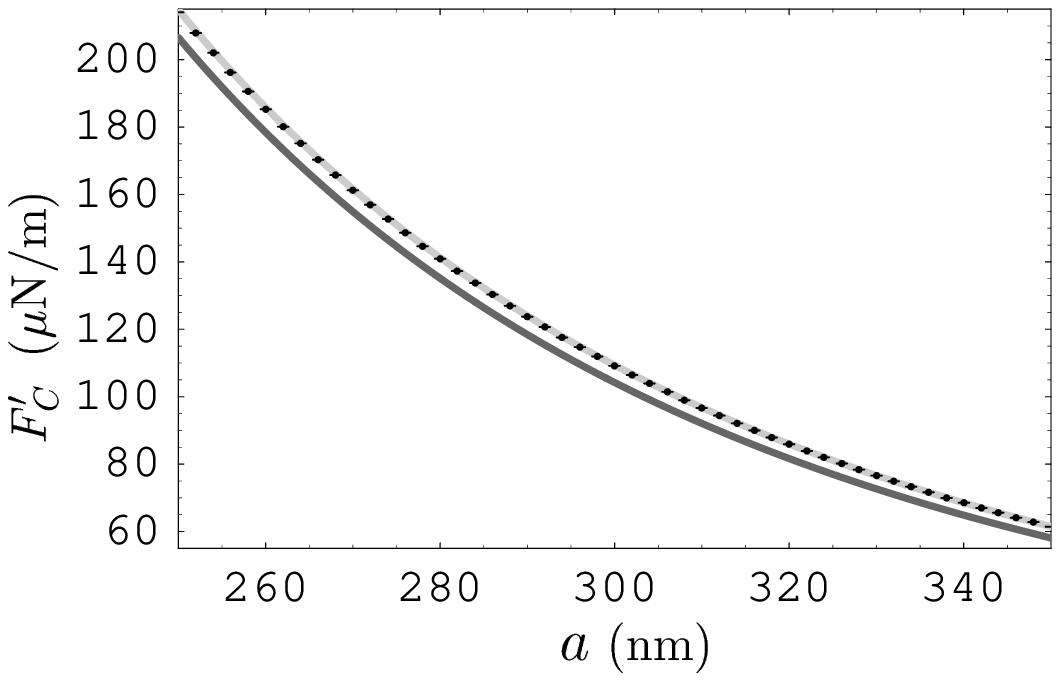}
}
\vspace*{-7cm}
\caption{\label{fg2}
The gradients of the Casimir force between Au surfaces measured
by means of a micromachined oscillator versus separation are
indicated as crosses. The dark- and light-gray bands show the
theoretical predictions using the Drude and plasma model
approaches, respectively.
}
\end{figure}

The second unexpected problem of the Lifshitz theory is that the
theoretical Casimir force between metallic test bodies was found
in drastic contradiction to the measurement data if the Drude
model (\ref{eq5}) is used at low frequencies. Alternatively,
the predictions of the Lifshitz theory were found in excellent
agreement with the measurement data when the dielectric permittivity
of Au was extrapolated to low frequencies by means of the plasma
model (\ref{eq7}). This situation is illustrated in
Fig.~\ref{fg2} where the predictions of the Lifshitz theory
for the gradient of the Casimir force acting between Au-coated
surfaces of a sphere and a plate are shown by the dark-gray and
light-gray bands when the dielectric permittivities (\ref{eq5})
and (\ref{eq7}), respectively, were used in computations.
The experimental data are shown as crosses whose arms indicate the
total experimental errors. As can be seen in Fig.~\ref{fg2},
the experimental data are in excellent agreement with the
Lifshitz theory using the dielectric permittivity (\ref{eq7})
(the so-called plasma model approach) and exclude the
predictions of the Lifshitz theory using the dielectric
permittivity (\ref{eq5}) (the Drude model approach).
Figure~\ref{fg2} is plotted by the results of the experiment
\cite{22} performed by means of a micromachined oscillator.
Similar results leading to the same conclusions were later
obtained by another experimental group by means of an atomic
force microscope \cite{46}.

Thus, the Lifshitz theory combined with the plasma model
was confirmed experimentally. It should be stressed that the
constraints on the Yukawa-type corrections, presented in Sec.~II,
are obtained from the measure of agreement of the data with this
theoretical approach.
Therefore any additional arguments concerning its
validity are highly desirable. First of all, we stress that the
difference between the dark-gray and
light-gray bands in Fig.~\ref{fg2} cannot be explained by the
presence of some hypothetical Yukawa-type force acting between a
sphere and a plate because of different dependences of these
quantities on separation.

It was further hypothesized \cite{47} that besides the Casimir
force there might be some additional force between the sphere and
the plate due to electrostatic patches caused by the grain
structure of the polycrystal Au coatings, dust and contaminants
on the surfaces. The force gradient due to electrostatic patches
is positive and leads to attraction. It was speculated \cite{47}
that when this force gradient is added to the theoretical
prediction of the Drude model approach it might bring the
resulting theoretical force in agreement with the experimental
data. Then an apparent disagreement of the data
with the Drude model approach would be explained.

In response to these arguments it was noted that the respective
patches should be of rather large size which is in direct
contradiction with the sizes of grains and the quality of
surfaces used in the experiments \cite{46}. However, in the
end of 2012 final experimental confirmation of the plasma
model approach was still missing providing possibility to
cast doubts on the reliability of constraints obtained from
the measure of agreement between experiment and theory.
Situation has been changed in 2013 when the first measurements
of the Casimir force between magnetic surfaces have been
performed.

\section{Measurements of the Casimir force between ferromagnetic
and corrugated surfaces lead to new constraints on
non-Newtonian gravity}

Although the Casimir interaction between ferromagnetic surfaces
was predicted in 1971 \cite{48}, it was experimentally
demonstrated for the first time quite recently \cite{49,49a}.
In Refs.~\cite{49,49a} the gradient of the Casimir force between
two Ni-coated surfaces of a sphere and a plate was measured by
means of an atomic force microscope. Measurement of the Casimir
interaction between two magnetic surfaces is of fundamental
importance
because it sheds additional light on the validity of different
approaches to the application of the Lifshitz theory and on the
role of possible background effects, such as patch potentials, in
theory-experiment comparison.

\begin{figure}[b]
\vspace*{-12cm}
\centerline{\hspace*{3cm}
\includegraphics{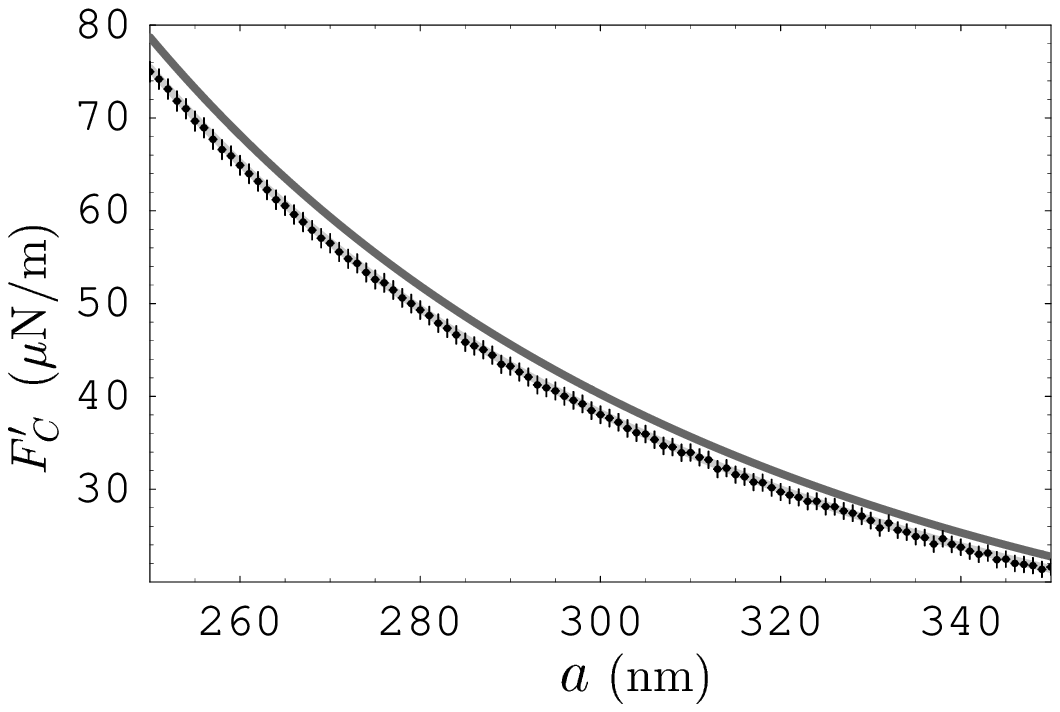}
}
\vspace*{-7cm}
\caption{\label{fg3}
The gradients of the Casimir force between Ni surfaces measured
by means of an atomic force microscope versus separation are
indicated as crosses. The dark- and light-gray bands show the
theoretical predictions using the Drude and plasma model
approaches, respectively.
}
\end{figure}

In Fig.~\ref{fg3} by the dark-gray and light-gray bands we show
theoretical predictions for the gradient of the Casimir force
between Ni-coated surfaces of a sphere and a plate calculated
using the Drude and plasma model approaches, respectively.
In the same figure, the experimental data with their total errors
are shown as crosses. As can be seen in Fig.~\ref{fg3},
the plasma model approach is again in excellent agreement with
the data whereas the Drude model approach is experimentally
excluded. In this respect Fig.~\ref{fg3} might be considered
as similar to Fig.~\ref{fg2} related to the case of nonmagnetic
(Au) surfaces. There is, however, the fundamental difference
between Figs.~\ref{fg3} and \ref{fg2}. The point is that for
magnetic surfaces (Fig.~\ref{fg3}) the Lifshitz theory combined
with the Drude model predicts larger force gradients than the
Lifshitz theory combined with the plasma model. This is exactly
the opposite
of that for nonmagnetic metals (Fig.~\ref{fg2}) where the Drude
model approach predicts smaller force gradients than the plasma
model approach. Thus, if one suggests that there is some
additional attraction due to patches (or some other background
effect) between Au surfaces, which brings the Drude model
approach in agreement with the measurement data, just this
addition would bring the data for Ni surfaces in disagreement
with both theoretical approaches.

One can conclude that measurements of the Casimir interaction
between magnetic surfaces confirm the smallness of possible
background effects in the aforementioned experiments using
an atomic force microscope and micromachined oscillator, so
that these effects do not influence on the comparison between
experiment and theory. It is also confirmed that the Lifshitz
theory using the plasma model at low frequencies correctly
describes the Casimir interaction between metallic test
bodies (in so doing the fundamental reasons behind this
conclusion await for further investigation).
The constraints on the Yukawa-type corrections to Newton's
gravitational law, following from measurements of the Casimir
force between magnetic surfaces, are obtained in Ref.~\cite{50}.
They are in qualitative agreement with the constraints
obtained in Refs.~\cite{22,23} (see lines 2 and 3 in
Fig.~\ref{fg1}),
but a bit weaker due to smaller density of Ni as compared
to Au. However, the main importance of the experiment with Ni
surfaces is that it has added confidence in all constraints
on the Yukawa-type corrections to Newtonian gravity obtained
from measurements of the Casimir interaction.

\begin{figure}[b]
\vspace*{-16cm}
\centerline{\hspace*{5cm}
\includegraphics{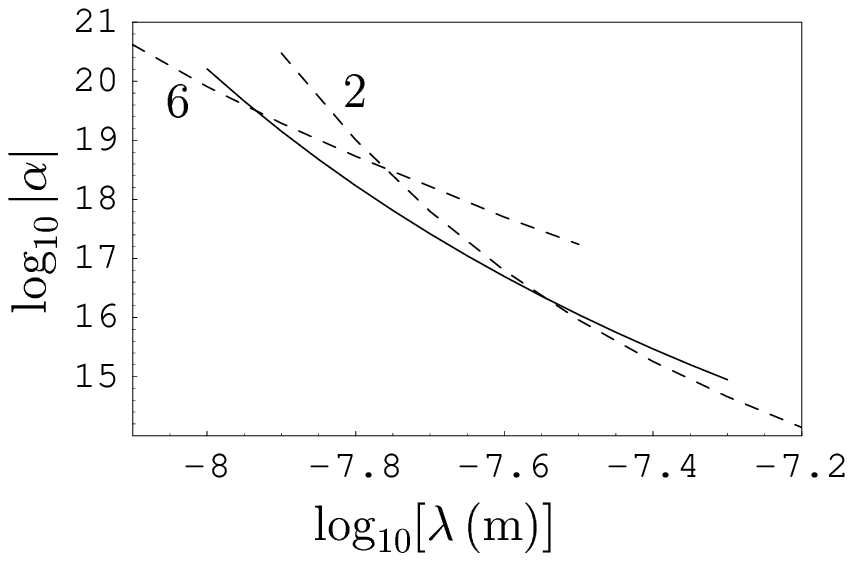}
}
\vspace*{-6cm}
\caption{\label{fg4}
Constraints on the  Yukawa-type corrections to
Newton's gravity from measurements of the Casimir force
between sinusoidally corrugated surfaces as compared with
other strongest constraints
(see  text for further discussion).
}
\end{figure}

Another recent experiment is on measurement of the Casimir
force between Au-coated sinusoidally corrugated surfaces of
a sphere and a plate \cite{51}. As opposite to
 Refs.~\cite{27,28}, here measurements were performed at various
 angles between corrugations. The corrugation periods on both
 test bodies were the same. The experimental results were found
 in good agreement with theoretical predictions using
 generalization of the Lifshitz theory for the case of nonplanar
 surfaces. The constraints on non-Newtonian gravity from this
 experiment were obtained in Ref.~\cite{50}. For this purpose
 the Yukawa-type force in the experimental configuration was
 calculated both exactly using Eq.~(\ref{eq2}) and
 approximately using the proximity force approximation with
 coinciding results \cite{52,53}. In Fig.~\ref{fg4} the
 solid line presents the most strong constraints on the
 Yukawa-type correction to Newtonian gravity which follow from
the experiment of Ref.~\cite{51} at the angle between
corrugations equal to $2.4^{\circ}$. In the same figure
the dashed lines 6 and 2 indicate the previously known
strongest constraints in this interaction range obtained from
measurements of the lateral Casimir force by means of an atomic
force microscope and the gradient of the Casimir force by means
of micromachined oscillator (in Fig.~\ref{fg1} these lines were
shown as the solid lines 6 and 2, respectively).
As is seen in Fig.~\ref{fg4}, the new constraints are stronger
than the previously known ones within the interaction region
from $\lambda=11.6\,$nm to $\lambda=29.2\,$nm. The maximum
strengthening by a factor 4 is achieved at $\lambda=17.2\,$nm.

\section{Conclusions and discussion}

In this paper we have discussed new constraints on the
Yukawa-type correction to Newton's gravitational law at short
separations obtained recently from measurements of the Casimir
interaction. These constraints were found from the measure of
agreement between the experimental data and the fundamental
theory of the van der Waals and Casimir forces developed by
Lifshitz. It was shown that the comparison of the
measurement data with this theory is a delicate problem.
According to the experimental results, the relaxation properties
of conduction electrons do not influence on the Casimir force
and should not be included in theory-experiment comparison.
Many persistent attempts to avoid this conclusion at the
cost of some background effects or possible inaccuracy in
calculations finally failed after recent demonstration of the
Casimir interaction between ferromagnetic surfaces.
At the moment the facts are known but the physical reasons
behind them invite further investigation.

Measurements of the Casimir force continue to be very
prospective for obtaining stronger constraints on the
Yukawa-type corrections to Newtonian gravity at short separations.
This was confirmed by the recent experiment with sinusoidally
corrugated boundary surfaces performed at different angles between
corrugations. The already achieved strengthening of the
constraints by a factor of 4 from this experiment can be further
improved due to some modifications in the measurement scheme.
This shows that measurements of the Casimir force at a
laboratory table continue to be an important source of
information on the border between quantum physics and gravitation
supplementary to information obtained from the accelerator
experiments,
astrophysics and cosmology.


\end{document}